\documentclass[aps,pra,twocolumn,showpacs,floatfix]{revtex4}
\usepackage{epsfig}
\usepackage{graphicx}
\usepackage{dcolumn}
\usepackage{amsthm,amsmath}
\usepackage{pstricks}
\usepackage{color}

\begin{document}

\title{Implementation of Z-vector method in the relativistic coupled cluster framework to calculate first order energy derivatives: Application to SrF molecule} 
\author{Sudip Sasmal\footnote{sk.sasmal@ncl.res.in}$^1$, Himadri Pathak$^1$, Malaya K. Nayak$^2$, Nayana Vaval$^1$, Sourav Pal$^1$ }
\affiliation{$^1$Electronic Structure Theory Group, Physical Chemistry Division, CSIR-National Chemical Laboratory, Pune, 411008, India}
\affiliation{$^2$Theoretical Chemistry Section, Chemistry Group, Bhabha Atomic Research Centre, Trombay, Mumbai 400085, India}
\begin{abstract}
The molecular dipole moment and magnetic hyperfine structure constant demand an accurate
wavefunction far from the nucleus and in near nuclear region, respectively.
We, therefore, employ the so-called Z-vector method in the domain of relativistic
coupled cluster theory to calculate the first order property of molecular
systems in their open-shell ground state configuration. The implemented method is applied to
calculate molecular dipole moment and parallel component of the magnetic
hyperfine structure constant of SrF molecule. The results of our calculation
are compared with the experimental and other available theoretically calculated values.
We are successful in achieving good accordance with the experimental results.
The result of our calculation of molecular dipole moment is in the accuracy of $\sim$  0.5 \%,
which is clearly an improvement over the previous calculation based on the expectation
value method in the four component coupled cluster framework 
[V. S. Prasannaa {\it et al,} Phys. Rev. A {\bf 90}, 052507 (2014)] and it is the best
calculated value till date.
Thus, it can be inferred that the Z-vector method can provide an accurate wavefunction in both near and far nuclear region, which is evident from our 
calculated results.
\end{abstract}
\maketitle
Theoretical physicists find it very challenging to calculate the spectroscopic properties of
atoms and molecules. The precise description of the spectroscopic properties
demands the wavefunction to be accurate both in the nuclear region and the region far from the nucleus. The
calculation of an accurate wavefunction involving heavy atoms and molecules needs to 
include the relativistic and electron correlation effects simultaneously, as these two effects are non-additive in nature
\cite{grant_book,lindgren_book}.
The best possible way to include the effects of relativity in a single determinantal theory is to solve the
Dirac-Hartree-Fock (DHF) Hamiltonian in its four component formalism. The DHF Hamiltonian converts
the complicated many electron problem into a sum of many one-electron problems by assuming an average 
electron-electron interaction. Therefore, the DHF Hamiltonian lacks the correlation of opposite spin electrons.
The missing electron correlation can be included by adding orthogonal space to the DHF wavefunction.
On the other hand, normal coupled cluster (NCC) \cite{cizek_1969,pal_1989,bartlett_2007} method is known to be 
the most elegant many-body theory to effectuate the dynamic part of the electron correlation.

The calculations of one electron response properties in 
the NCC framework, can either be done by taking
expectation value of the desired property operator or as a derivative of energy.
These two approaches are not same as the NCC is by nature non-variational.
In fact, the first order derivative of energy is the 
corresponding expectation value plus some additional terms, which makes the derivative approach closer to the full
configuration interaction (FCI) property value. It is worth to mention that the expectation value
approach in the NCC leads to a nonterminating series and any truncation scheme introduces an additional error
\cite{kutzelnigg_1991}.

In general, the energy is a function of both the determinantal coefficients ($C_D$) in the expansion of the 
many electron correlated wavefunction and the molecular orbital coefficients ($C_M$) for a fixed nuclear geometry.
The first order energy derivative in NCC can be written as
\begin{eqnarray}
\frac{\delta E[C_D(\lambda),C_M(\lambda)]}{\delta \lambda} = \frac{\delta E}{\delta C_D} \frac{\delta C_D}{\delta \lambda}
+ \frac{\delta E}{\delta C_M} \frac{\delta C_M}{\delta \lambda}. \nonumber
\end{eqnarray}
Thus, for the calculation of energy derivative in NCC framework, it is, therefore, necessary to calculate the derivative
of energy with respect the determinantal coefficients as well as the molecular orbital coefficients. It further requires the
derivative of the determinantal coefficients and molecular orbital coefficients with respect to the external field of 
perturbation. However, Bartlett and co-workers \cite{bartlett_1986} have shown that these derivative terms 
can be transformed
into a single linear equation by using Z-vector method. The advantage of Z-vector method 
\cite{schaefer_zvec,bartlett_zvec} is that for the
calculation of several properties, one needs to solve a single linear equation instead of solving equations
for each external perturbation field of interest. The detailed diagrammatic of Z-vector method in NCC
framework is given in Ref. \cite{bartlett_zvec}.

In this rapid communication, we have shown that the Z-vector method in the NCC framework within its four component description can generate
an accurate wavefunction in the near nuclear region as well as in the region far from the nucleus. To justify our argument,
we have compared
magnetic hyperfine structure constant (HFS) and molecular dipole moment of SrF with the experimental values
as the calculation of these properties need an accurate wavefunction in the near nuclear region and  
the region far from the nucleus, respectively.
The reason for choosing SrF molecule is as follows:
The knowledge of long range
dipole-dipole interaction is very important to produce ultracold molecules in optical lattice
\cite{trefzger_2011}. SrF molecule can be cooled by laser spectroscopy \cite{shuman_2010} and thus
it can be used for high precession spectroscopy \cite{mathavan_2014,berg_2014}.
Currently an experimental search for parity violation using SrF is in progress \cite{berg_p_violation}.
Therefore, detailed knowledge
of the spectroscopic properties like dipole moment and magnetic HFS is very important to interpret the
experimental findings.

The Many body time-independent Dirac-Coulomb Hamiltonian is given by
\begin{eqnarray}
  H = \sum_i \Big [-c\vec{\alpha}_i\cdot\vec{\nabla}_i + (\beta -I)c^2 + \sum_A V_A^{nuc}(r_i) +
\sum_{j>i} \frac{1}{r_{ij}} \Big ],
\end{eqnarray}
where $c$ is the speed of light, {\bf$\alpha$} and $\beta$ are the usual Dirac matrices,
$I$ is the 4$\times$4 identity matrix.
$V_A^{nuc}(r_i)$ is the nuclear potential term of nucleus A.

The dynamic part of the electron correlation is included using the coupled cluster method.
The wave function in the coupled cluster method is defined as 
\begin{eqnarray}
 | \Psi_{cc} \rangle = e^{T} | \Phi_0 \rangle
\end{eqnarray}
where, $|\Phi_0\rangle$ is the ground state single determinant wavefunction and $T$ is cluster operator which is given
by
\begin{eqnarray}
 T=T_1+T_2+...+T_N=\sum_n^N T_n
\end{eqnarray}
with
\begin{eqnarray}
 T_m= \frac{1}{(m!)^2} \sum_{ij..ab..} t_{ij..}^{ab..}{a_a^{\dagger}a_b^{\dagger} .. ..a_j a_i}
\end{eqnarray}
i,j(a,b) are the hole(particle) indices and $t_{ij..}^{ab..}$ are the cluster amplitudes corresponding 
to the cluster operator $T_m$.
The equations for n-body cluster amplitudes and correlation energy are given by
\begin{eqnarray}
 \langle \Phi_{i..}^{a..} | (H_Ne^T)_C | \Phi_0 \rangle = 0
\label{cc_amplitudes}
\end{eqnarray}
\begin{eqnarray}
 \langle \Phi_0 | (H_Ne^T)_C | \Phi_0 \rangle = E^{corr}
\label{cc_energy}
\end{eqnarray}
where $H_N$ is the normal ordered Hamiltonian and the subscript $C$ indicates only the connected terms
in the contraction between $H_N$ and $T$. The connectedness ensures size-extensivity.
Once the cluster amplitude equation (\ref{cc_amplitudes}) is solved, the correlation
energy can be obtained from equation \ref{cc_energy}.
The coupled cluster energy is a function of both the determinantal coefficients ($C_D$) and the molecular orbital coefficients ($C_M$).
Therefore, the calculation of coupled cluster energy derivative
need both the derivative of $C_Ds$ and $C_Ms$ with respect to external field of perturbation. However,
the equations involving derivative of $C_Ds$ and $C_Ms$ are linear equations. Thus, one needs to solve the
linear equations for each external field perturbation of interest. This can be avoided with the introduction of 
an antisymmetrized de-excitation operator, $\Lambda$, where the equations for the amplitudes of $\Lambda$ is perturbation
independent. Therefore, the solution of one linear equation is required instead of solving for each external perturbation.
The second quantized form of the perturbation independent operator, $\Lambda$ is given by
\begin{eqnarray}
 \Lambda=\Lambda_1+\Lambda_2+...+\Lambda_N=\sum_n^N \Lambda_n
\end{eqnarray}
where
\begin{eqnarray}
 \Lambda_m= \frac{1}{(m!)^2} \sum_{ij..ab..} \lambda_{ab..}^{ij..}{a_i^{\dagger}a_j^{\dagger} .. ..a_b a_a}
\end{eqnarray}
where i,j(a,b) are the hole(particle) indices and $\lambda_{ab..}^{ij..}$ are the cluster amplitudes corresponding 
to the cluster operator $\Lambda_m$.
The detailed description of $\Lambda$ operator and $\Lambda$ amplitude equation is given in reference \cite{bartlett_zvec}.
The working $\Lambda$ amplitude equation is given by
\begin{eqnarray}
  \langle \Phi_0 |& [\Lambda (H_Ne^T)_C]_C | \Phi_{i..}^{a..} \rangle + \langle \Phi_0 | (H_Ne^T)_C | \Phi_{int} \rangle \nonumber\\
 &  \langle \Phi_{int} | \Lambda | \Phi_{i..}^{a..} \rangle + \langle \Phi_0 | (H_Ne^T)_C | \Phi_{i..}^{a..} \rangle = 0
 \end{eqnarray}
where $\Phi_{int}$ is the determinant corresponding to the intermediate excitation between $\Phi_0$
and $\Phi_{i..}^{a..}$.
In the coupled cluster single and double (CCSD) model, $\Lambda$ becomes, $\Lambda=\Lambda_1+\Lambda_2$.
The explicit equations for the amplitudes of $\Lambda_1$ and $\Lambda_2$ operators are
\begin{eqnarray}
  \langle \Phi_0 |& [\Lambda (H_Ne^T)_C]_C | \Phi_{i}^{a} \rangle + \langle \Phi_0 | (H_Ne^T)_C | \Phi_{i}^{a} \rangle = 0,
\end{eqnarray}
\begin{eqnarray}
  \langle \Phi_0 |& [\Lambda (H_Ne^T)_C]_C | \Phi_{ij}^{ab} \rangle + \langle \Phi_0 | (H_Ne^T)_C | \Phi_{i}^{a} \rangle \nonumber\\
 &  \langle \Phi_{i}^{a} | \Lambda | \Phi_{ij}^{ab} \rangle + \langle \Phi_0 | (H_Ne^T)_C | \Phi_{ij}^{ab} \rangle = 0.
\label{lambda_2}
\end{eqnarray}
\begin{figure}[b]
\centering
\begin{center}
\includegraphics[height=3 cm, width=3.0 cm]{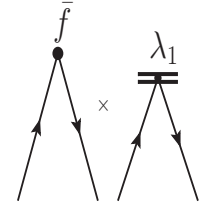}
\caption {Disconnected yet linked diagram in $\Lambda_2$ equation.}
\label{disconnected}
\end{center}  
\end{figure}
It is interesting to note that the term $\langle \Phi_0 | (H_Ne^T)_C | \Phi_{i}^{a} \rangle
\langle \Phi_{i}^{a} | \Lambda | \Phi_{ij}^{ab} \rangle$ of equation \ref{lambda_2} yields one disconnected
diagram, which is given in Fig. \ref{disconnected}.
The said diagram is not of the type of closed with disconnected part.
This ensures that the energy derivative is linked and thus
size extensive.
The equation for energy derivative can be written as
\begin{eqnarray}
\Delta E' = \langle \Phi_0 | (O_Ne^T)_C | \Phi_0 \rangle + \langle \Phi_0 | [\Lambda (Oe^T)_C]_C | \Phi_0 \rangle
\end{eqnarray}
where $O_N$ is the derivative of normal ordered perturbed Hamiltonian with respect to external field of perturbation.

The molecular dipole moment of a heavy diatomic molecule arises due to the fact that the nuclear charge of the two atoms
are not same and thus the electron density is not evenly distributed around the two nucleus. In the Born-Oppenheimer
approximation, we can separate out the nuclear and electronic contribution from the total contribution. The dipole moment
operator is given by
\begin{eqnarray}
\vec{\mu} = - \sum_i \vec{r}_i + \sum_A Z_A\vec{r}_A
\end{eqnarray}
where $i$ stands for the electron and $Z_A$ is the atomic number of nucleus $A$. The first term of the above equation
is the electronic contribution and the second term is the nuclear contribution.

The magnetic HFS arises due to the interaction of nuclear magnetic dipole moment with the magnetic moment
of electrons. Thus, the magnetic HFS is the sum of one body interactions from the point of view of electronic  
structure theory \cite{lindgren_book}. The magnetic vector potential $(\vec{A})$ at a distance $\vec{r}$ is given by
\begin{equation}
 \vec{A}=\frac{\vec{\mu}_k \times \vec{r}}{r^3},
\end{equation}
where $\vec{\mu}_k$ is the magnetic moment of the nucleus $K$. The perturbed HFS Hamiltonian
of an atom due to $\vec{A}$ in the Dirac theory is given by
$H_{hyp}= \sum_i^n \vec{\alpha}_i \cdot \vec{A_i}$,
where $n$ is the total no of electrons and $\alpha_i$ denotes the Dirac $\alpha$
matrices for the i$^{th}$ electron. The z projection (along the molecular axis) of the expectation value 
of the corresponding
perturbed HFS Hamiltonian gives the parallel magnetic HFS constant, $A_{\|}$, which is given by
\begin{eqnarray}
A_{\|}= \frac{\vec{\mu_k}}{I\Omega} \cdot \langle \Psi_{\Omega} | \sum_i^n
\left( \frac{\vec{\alpha}_i \times \vec{r}_i}{r_i^3} \right)_{z} | \Psi_{\Omega}  \rangle,
\end{eqnarray}
where $I$ is the nuclear spin quantum number and $\Omega$ represents the z component
(along molecular axis) of the total angular momentum of the diatomic molecule.

The N electron ground and excited determinants are constructed with the one electron spinors, those
are the solutions of Dirac-Hartree-Fock equation. The DIRAC10 \cite{dirac10} program package is used to solve
the Dirac-Fock
equation and to obtain the matrix elements required for property calculations. Gaussian charge distribution is considered as the nuclear
model where the nuclear parameters \cite{visscher_1997} are taken as default values in DIRAC10.
Large and small component
basis functions are generated by applying restricted kinetic balance (RKB) \cite{dyall_2006} in which
basis functions
are represented in scalar basis and unphysical solutions are removed by diagonalizing
the free particle Hamiltonian. This generates the electronic and positronic solution in 1:1 manner.
We have done five different sets of calculation using five different basis sets for Sr and F. These are
cc-pVDZ, aug-cc-pVTZ, aug-cc-pCVTZ, d-aug-cc-pCVTZ and aug-cc-pCVQZ for F atom \cite{dunning_f} and
dyall.v2z, dyall.v3z, dyall.cv3z, d-aug-dyall.cv3z and dyall.cv4z for Sr atom \cite{dyall_sr}.
Both large and small component basis are taken
in uncontracted form. None of the electrons are frozen in our correlation calculation and the virtual
orbitals whose energies are greater than a certain threshold are not considered as
the high energy virtual orbitals contribute less in correlation calculation.

We have taken the following strategies to code the Z-vector method. First, the one electron and
two electron matrix elements are obtained from DIRAC10 package \cite{dirac10}. Then we have solved
the NCC part i.e., the $T_1$ and $T_2$ amplitude equations. This is followed by the construction of
different types of $\bar{H}$ ($\bar{H}=(He^T)_C$).
After that $\bar{H}$ vertices are contracted with one $\Lambda_1$ or $\Lambda_2$ vertex to construct
the $\Lambda_1$ and $\Lambda_2$ amplitude equations. At the end $T_1$, $T_2$, $\Lambda_1$ and $\Lambda_2$
amplitudes are contracted with property integrals to get corresponding property value. To solve the
$T_1$ and $T_2$ amplitudes and to construct the $\bar{H}$, we have used a recursive intermediate 
factorization of diagrams as described by Bartlett and coworkers \cite{kucharski_1991}. This saves
enormous computational cost.
\begin{table*}[ht]
\caption{ Cutoff used and correlation energy of the ground state of SrF in different basis sets }
\begin{ruledtabular}
{%
\newcommand{\mc}[3]{\multicolumn{#1}{#2}{#3}}
\begin{center}
\begin{tabular}{lrrrrrr}
\mc{3}{c}{Basis} & Cutoff  & Spinor & \mc{2}{c}{Correlation Energy (a.u.)}\\
\cline{1-3} \cline{6-7}\\
Name & \mc{1}{c}{Sr} & \mc{1}{c}{F} & (a.u.) & & MBPT(2) & CCSD \\
\hline
A & dyall.v2z & cc-pVDZ &   & 298 & -1.528622377 & -1.474175309 \\
B & dyall.v3z & aug-cc-pVTZ & 500 & 366 & -1.292096889 & -1.256789931 \\
C & dyall.cv3z & aug-cc-pCVTZ & 500 & 436 & -1.838672666 & -1.775710294 \\
D & d-aug-dyall.cv3z & d-aug-cc-pCVTZ & 100 & 596 & -1.681326043 & -1.621603448 \\
E & dyall.cv4z & aug-cc-pCVQZ & 50 & 520 & -1.454995437 & -1.402836367
\end{tabular}
\end{center}
}%
\end{ruledtabular}
\label{basis}
\end{table*}
To debug this code, we benchmarked our correlation energy with the results obtained from
DIRAC10 with same basis, same convergence criteria and using same direct inversion in
the iterative subspace (DIIS). We have achieved 7 to 8 decimal place agreement with DIRAC10 for correlation
energy independent of the choice of molecules as well as of the basis sets. The discrepancy beyond this
limit could be due to the use of cutoff in storing of the intermediate diagrams or the use of different
convergence algorithm. The $H$ and $\bar{H}$ matrix elements are stored by setting a cutoff of 10$^{-12}$ to save
storage requirement as the contribution of the two body matrix elements beyond that limit is negligible. The tolerance 
used for the convergence of both $T$ and $\Lambda$ amplitudes is 10$^{-9}$. We have used the experimental
bond length (2.075 \AA) of SrF \cite{bond_len} in all the calculations.

In Table \ref{basis}, we present the sets of basis used in our calculations and each combination
is denoted by an English alphabet letter. The fourth and fifth column of Table \ref{basis}
represent the cutoff used and the number of spinor generated using that cutoff for correlation
calculation, respectively. We also compiled the correlation energy of SrF obtained from CCSD and 
second-order many body perturbation theory (MBPT(2)), which uses a first-order perturbed
wavefunction.

In Table \ref{dipole}, we present the molecular dipole moment in Debye of the ground state of SrF
molecule in five
different basis sets. The experimental value \cite{ernst_1985} is also presented in the same table
for comparison.
It is clear from the table that with increase in the number of basis function the dipole moment
converges towards the experimental value. This is expected as more basis functions generate more
correlation space and thereby improve the dipole moment. In particular, our calculated dipole
moment in basis E is very close ($\sim$ 0.5 \%) to the experimental value.
\begin{table}[h]
\caption{ Molecular dipole moment (in Debye) of the ground state of SrF}
\begin{ruledtabular}
\begin{center}
\begin{tabular}{lrr}
Basis & Z-vector & Experiment \cite{ernst_1985}\\
\hline 
A  & 3.0158 & \\
B  & 3.3898 &  \\
C  & 3.4023 & 3.4676(10) \\
D  & 3.4376 & \\
E  & 3.4504 & \\
\end{tabular}
\end{center}
\end{ruledtabular}
\label{dipole}
\end{table}
\begin{table}[ht]
\caption{ Comparison of molecular dipole moment of the ground state of SrF in different methods}
\begin{ruledtabular}
\newcommand{\mc}[3]{\multicolumn{#1}{#2}{#3}}
\begin{center}
\begin{tabular}{lrr}
Method & Reference & Dipole (D)\\
\hline
Ionic model & Torring {\it et al.} \cite{torring_1984} & 3.67\\
SCF & Langhoff {\it et al.} \cite{langhoff_1986} & 2.579\\
CPF & Langhoff {\it et al.} \cite{langhoff_1986} & 3.199\\
CISD & Langhoff {\it et al.} \cite{langhoff_1986} & 2.523\\
EPM & Mestdagh {\it et al.} \cite{mestdagh_1991} & 3.6\\
HF (finite difference) & Kobus {\it et al.} \cite{kobus_2000} & 2.5759\\
CCSD & Prasannaa {\it et al.} \cite{prasannaa_2014} & 3.41\\
Z-vector & This work(E) & 3.4504 \\
Expt. & Ernst {\it et al.} \cite{ernst_1985} & 3.4676(10)
\end{tabular}
\end{center}
\end{ruledtabular}
\label{comparison}
\end{table}
The results obtained for the dipole moment of the ground state of SrF by other methods and experiment are compiled
in Table \ref{comparison}. The dipole moment of SrF was first calculated by Torring {\it et al,}
\cite{torring_1984}
by using an ionic model and they got a value of 3.67 D. Langhoff {\it et al,} \cite{langhoff_1986}
performed the first {\it ab initio} calculation
of the dipole moment of SrF by using Slater type of basis function. They reported the dipole
moment using three different methods, i.e., self consistent field (SCF), configuration interaction
in single and double approximation (CISD) and the coupled pair function (CPF) method. Among them,
CISD method is not size extensive while CPF is, thus CPF approach gives
better agreement with experiment. However, Langhoff {\it et al,} did not consider the relativistic
motion of electrons. Mestdagh {\it et al,} \cite{mestdagh_1991} used electrostatic polarization model and got 3.6 D
as a molecular dipole moment of SrF. Kobus {\it et al,} \cite{kobus_2000} obtained a dipole moment of 2.5759 D by using
finite difference method in the Hartree-Fock (HF) level. The first relativistic
calculation of dipole moment of SrF in the CCSD model was calculated by Prasannaa {\it et al,}
\cite{prasannaa_2014} by taking expectation value of the corresponding operator.
The expectation value framework leads to a connected yet nonterminating series. Prasannaa {\it et al,}
took only the linear terms in the property calculations using CCSD wavefunction and got 3.41 D as a result.
Our four component Z-vector calculation gives a result of 3.4504 D
by using dyall.cv4z basis for Sr \cite{dyall_sr} and aug-cc-pCVQZ basis for F \cite{dyall_sr} (basis E)
and this result shows the best agreement with the experiment so far.
\begin{table*}[ht]
\caption{Parallel ($A_{\|}$) magnetic hyperfine structure constant of the ground state of SrF in MHz }
\begin{ruledtabular}
\newcommand{\mc}[3]{\multicolumn{#1}{#2}{#3}}
\begin{center}
\begin{tabular}{lrrrr}
 & \mc{2}{c}{$^{87}$Sr} & \mc{2}{c}{$^{19}$F}\\
\cline{2-3} \cline{4-5}\\
Basis & Z-vector & Experiment \cite{weltner_1983} & Z-vector & Experiment \cite{weltner_1983} \\
\hline
A & 546.08 &  & 121.93 & \\
B & 558.96 &  & 118.70 & \\
C & 566.62 & 591(3) & 119.64 & 126(3) \\
D & 561.25 &  & 116.35 & \\
E & 559.65 &  & 117.74 & \\
\end{tabular}
\end{center}
\end{ruledtabular}
\label{hfs}
\end{table*}

In Table \ref{hfs}, we present the parallel component of the magnetic HFS constant of $^{87}$Sr and $^{19}$F 
of the ground state of SrF molecule. We also present the experimental value \cite{weltner_1983} of those in the 
same table for comparison.
Our calculated result using Z-vector method show good agreement with the experimental result.
The highest and lowest deviation from experimental values for parallel magnetic HFS constant
of $^{87}$Sr atom are for the A ($\sim$ 45 MHz) and C ($\sim$ 24 MHz) basis respectively.
For the parallel magnetic HFS constant of $^{19}$F, the maximum and minimum deviation occur
for the basis D ($\sim$ 10 MHz) and A ($\sim$ 4 MHz) basis.

The calculated magnetic HFS constant values are in good agreement with the sophisticated experiment
but the extent of accuracy is not so in comparison to that of the calculated dipole moment values. This could
possibly be due to the fact that 
as we proceed from basis A to E, we have added extra Gaussian type
orbitals (GTOs) of higher angular momentum in addition to lower angular momentum.
As the higher angular momentum GTO shifts the electron density towards the outer region,
the addition of
higher angular momentum GTO improves the outer region much better than the inner region of the molecular
wavefunction. This is why as we go from basis A to E, our molecular dipole moment value matches
more closely than the magnetic HFS values with the experimental results.\\

We have successfully implemented Z-vector method in the relativistic NCC domain using 4-component wave function to calculate the 
first order energy derivatives. We applied this method to calculate the molecular dipole
moment and parallel magnetic HFS constant of SrF molecule.
The results from our calculations are in good agreement with the experimental values.
Therefore, we can conclude that Z-vector method in the relativistic framework can produce an accurate wavefunction in the
near nuclear region as well as far from the nucleus.

Authors acknowledge a grant from CSIR XII$^{th}$ Five Year
Plan project on Multi-Scale Simulations of Material (MSM) and the resources
of the Center of Excellence in Scientific Computing at CSIR-NCL.
S.S. and H.P acknowledge the CSIR for their fellowship. 


\begin{thebibliography}{}

\bibitem{grant_book}

I. P. Grant, {\it Relativistic Quantum Theory of Atoms and Molecules: Theory and Computation} (Springer,
New York, 2010).

\bibitem{lindgren_book}

I. Lindgren and J. Morrison, {\it Atomic Many-Body Theory } (Springer-Verlag, New York, 1985).

\bibitem{cizek_1969}

J. \v{C}\'{i}\v{z}ek, Adv. Chem. Phys. {\bf14}, 35 (1969).

\bibitem{pal_1989}

D. Mukherjee and S. Pal, Adv. Quantum Chem. {\bf20}, 291 (1989).

\bibitem{bartlett_2007}

R. J. Bartlett and M. Musia\l{}, Rev. Mod. Phys. {\bf79}, 291 (2007).

\bibitem{kutzelnigg_1991}

W. Kutzelnigg, Theor. Chim. Acta {\bf80}, 349 (1991).

\bibitem{bartlett_1986}

G. Fitzgerald, R. J. Harrison and R. J. Bartlett, J. Chem. Phys. {\bf85}, 5143 (1986).

\bibitem{schaefer_zvec}

N. C. Handy and H. F. Schaefer III, J. Chem. Phys. {\bf81}, 5031 (1981).

\bibitem{bartlett_zvec}

E. A. Salter, G. W. Trucks, and R. J. Bartlett, J. Chem. Phys. {\bf90}, 1752 (1989).

\bibitem{trefzger_2011}

C. Trefzger, C. Menotti, B. Capogrosso-Sansone, and M. Lewenstein, J. Phys. B: At. Mol.
Opt. Phys. {\bf44}, 193001 (2011).

\bibitem{shuman_2010}

E. S. Shuman, S. F. Barry, and D. DeMille, Nature {\bf467}, 820 (2010).

\bibitem{mathavan_2014}

S. C. Mathavan, C. Meinema, J. E. van den Berg, H. L. Bethlem, and S. Hoekstra,
``Deceleration, cooling and trapping of SrF molecules for precision spectroscopy,''
46th Conference of the European Groups on Atomic Systems, 2014,
(http://egas46.univ-lille1.fr/docs/EGAS46\_book\_of\_abstracts.pdf, Page 54)

\bibitem{berg_2014}
J. E. van den Berg, S. C. Mathavan, C. Meinema, J. Nauta, T. H. Nijbroek, K. Jungmann,
H. L. Bethlem, and S. Hoekstra, J. Mol. Spectrosc. {\bf300}, 2225 (2014).

\bibitem{berg_p_violation}

J. van den Berg, ``Using cold molecules to detect molecular parity violation,'' SSP2012, Groningen,
https://www.kvi.nl/ssp2012/material/107-berg,\_van\_den/slides/107-1-praatjeSSP2012.pdf

\bibitem{dirac10}

DIRAC, a relativistic ab initio electronic structure program, Release {DIRAC10} (2010),
written by T. Saue, L. Visscher and H. J. {\relax Aa}. Jensen,
with contributions from R. Bast, K. G. Dyall,
U. Ekstr{\"o}m, E. Eliav, T. Enevoldsen, T. Fleig, A. S. P. Gomes,
J. Henriksson, M. Ilia{\v{s}}, Ch. R. Jacob, S. Knecht, H. S. Nataraj, P. Norman,
J. Olsen, M. Pernpointner, K. Ruud, B. Schimmelpfennig, J. Sikkema,
A. Thorvaldsen, J. Thyssen, S. Villaume, and S. Yamamoto.
(see http://dirac.chem.vu.nl).

\bibitem{visscher_1997}

L. Visscher and K. G. Dyall, At. Data Nucl. Data Tabl. {\bf67}, 207 (1997)

\bibitem{dyall_2006}
K. G. Dyall and K. Faegri, Jr., {\it Introduction to Relativistic Quantum Chemistry}
(Oxford University Press, New York, 2006).

\bibitem{dunning_f}

T. H. Dunning, Jr. J. Chem. Phys. {\bf90}, 1007 (1989).

\bibitem{dyall_sr}

K.G. Dyall, J. Phys. Chem. A {\bf113}, 12638 (2009).

\bibitem{kucharski_1991}

S. A. Kucharski and R. J. Bartlett, Theor. Chim. Acta {\bf80}, 387 (1991).

\bibitem{bond_len}

{\it Molecular Spectra and Molecular Structure, IV. Constants of
Diatomic Molecules}, edited by K. P. Huber and G. Herzberg
(Van Nostrand Reinhold, New York, 1979).

\bibitem{ernst_1985}

W. E. Ernst, J. Kandler, S. Kindt, and T. Torring, Chem. Phys. Lett. {\bf113}, 351 (1985).

\bibitem{torring_1984}

T. Torring, W. E. Ernst, and S. Kindt, J. Chem. Phys. {\bf81}, 4614 (1984).

\bibitem{langhoff_1986}

S. R. Langhoff, C. W. Bauschlicher, Jr., H. Partridge, and R. Ahlrichs, J. Chem. Phys. {\bf84}, 5025 (1986).

\bibitem{mestdagh_1991}

J. M. Mestdagh and J. P. Visticot, Chem. Phys. {\bf155}, 79 (1991).

\bibitem{kobus_2000}

J. Kobus, D. Moncrieff, and S.Wilson, Phys. Rev. A {\bf62}, 062503 (2000).

\bibitem{prasannaa_2014}

V. S. Prasannaa, M. Abe, and B. P. Das, Phys. Rev. A {\bf90}, 052507 (2014).

\bibitem{weltner_1983}

W. Weltner, {\it Magnetic Atoms and Molecules} (Dover Publications Inc., New York, 1983).


\end{thebibliography}
\end{document}